\documentclass[aps,pre,twocolumn,groupedaddress,superscriptaddress,showpacs,showkeys]{revtex4-1}

\usepackage{amssymb}
\usepackage[figuresright]{rotating}

\begin{document}

\title{Watersheds and Explosive percolation}

\author{Hans J. Herrmann}
    \email{hans@ifb.baug.ethz.ch}
    \affiliation{Computational Physics for Engineering Materials, IfB, ETH Zurich, Schafmattstr. 6, 8093 Zurich, Switzerland}

\author{Nuno A. M. Ara\'ujo}
    \email{nuno@ethz.ch}
    \affiliation{Departamento de F\'isica, Universidade Federal do Cear\'a, Campus do Pici, 60451-970 Fortaleza, Cear\'a, Brazil}

\begin{abstract}
The recent work by Achlioptas, D'Souza, and Spencer opened up the possibility of obtaining a discontinuous (explosive) percolation transition by changing the stochastic rule of bond occupation. 
Despite the active research on this subject, several questions still remain open about the leading mechanism and the properties of the system.
We review the \textit{largest cluster} and the \textit{Gaussian} models recently introduced.
We show that, to obtain a discontinuous transition it is solely necessary to control the size of the largest cluster, suppressing the growth of a cluster differing significantly, in size, from the average one. 
As expected for a discontinuous transition, a Gaussian cluster-size distribution and compact clusters are obtained.
The surface of the clusters is fractal, with the same fractal dimension of the watershed line.

\end{abstract}


\maketitle

\section{Introduction}\label{sec::intro}

  Percolation, the paradigm for random connectivity, has since Hammersley \cite{Broadbent57} been one of the most often applied statistical models \cite{Stauffer94,Sahimi94}.
  Its phase transition being related to magnetic models \cite{Fortuin72} is in all dimensions one of the most robust continuous transitions known.
  This explains the enormous excitement generated by the recent work by Achlioptas, D'Souza, and Spencer \cite{Achlioptas09} describing a stochastic rule apparently yielding a discontinuous percolation transition on a fully connected graph.
  A discontinuous percolation transition is observed when the growth of the largest cluster is systematically suppressed \cite{Araujo10}, promoting the formation of several large components that eventually merge in an explosive way \cite{Friedman09}.
  Several aggregation models, based on percolation, have been developed to achieve this change in the nature of the transition \cite{Achlioptas09,Moreira10,Cho10,DSouza10,Araujo10,Manna10,Chen11}.
  These models are generally classified as {\it explosive percolation}, the name given in the original work that triggered the field \cite{Achlioptas09}.
  In that work, a {\it best-of-two} product rule is proposed where, at each iteration, two unoccupied bonds are randomly selected but only the one which minimizes the product of the mass of the clusters, connected with the bond, is occupied.
  This work was originally studied by Achlioptas {\it et al.} \cite{Achlioptas09} on a fully connected graph and analyzed in detail by Friedman and Landsberg \cite{Friedman09}.
  Ziff reported simulations on a regular square lattice \cite{Ziff09,Ziff10}, while Radicchi and Fortunato \cite{Radicchi09,Radicchi10} and Cho {\it et al.} \cite{Cho09} on scale-free networks.
  However, reported results of finite-size studies and size distributions are not consistent with a discontinuous transition.
  For example, Ziff \cite{Ziff10}, Radicchi and Fortunato \cite{Radicchi10}, as well as Cho {\it et al.} \cite{Cho10} found a power-law cluster-size distribution with an exponent close to two.
  Although, different from the exponent of classical percolation, the sole fact of finding a power law is untypical for discontinuous transitions.
  Also unusual for a discontinuous transition is that the clusters are fractal, as we found happens for the Achlioptas rule, from the behavior of the order parameter with the system size \cite{Ziff10,Radicchi10}.
  Since then, various rules have been devised \cite{Araujo10,Friedman09,Moreira10,Manna10,DSouza10}, all attempting a discontinuous transition towards an infinite cluster.
  In all proposed models one tries to keep the clusters of similar size and, for random graphs, the internal bonds of clusters should also be additionally suppressed \cite{Achlioptas09,Moreira10}.
  For example, Moreira \textit{et al.} \cite{Moreira10} have proposed a Hamiltonian formalism which provides a clear connection between equilibrium statistical mechanics and explosive percolation.
  They have shown that, for obtaining a discontinuous transition the size of the growing clusters should be kept approximately the same and, on random graphs, merging bonds (connecting different clusters) should dominate over the redundant ones (connecting sites in the same cluster).
  In this manuscript we review the work introduced by Ara\'ujo and Herrmann \cite{Araujo10} where two models are proposed yielding clear discontinuous transitions: the \textit{largest cluster} and the \textit{Gaussian} models.
  In the latter, at the percolation threshold, a bimodal cluster size distribution is found consistent with the nature of the transition.
  For both models, the cluster perimeters are fractal with a fractal dimension of $1.23\pm0.03$, similar to the one observed for watersheds \cite{Fehr09,Fehr11} and other models \cite{Porto99,Andrade09,Oliveira11}.

\begin{figure*}
\begin{center}
  \begin{tabular}{cc}
    \textit{Largest cluster model} & \textit{Gaussian model}\\
    \includegraphics[width=0.4\textwidth]{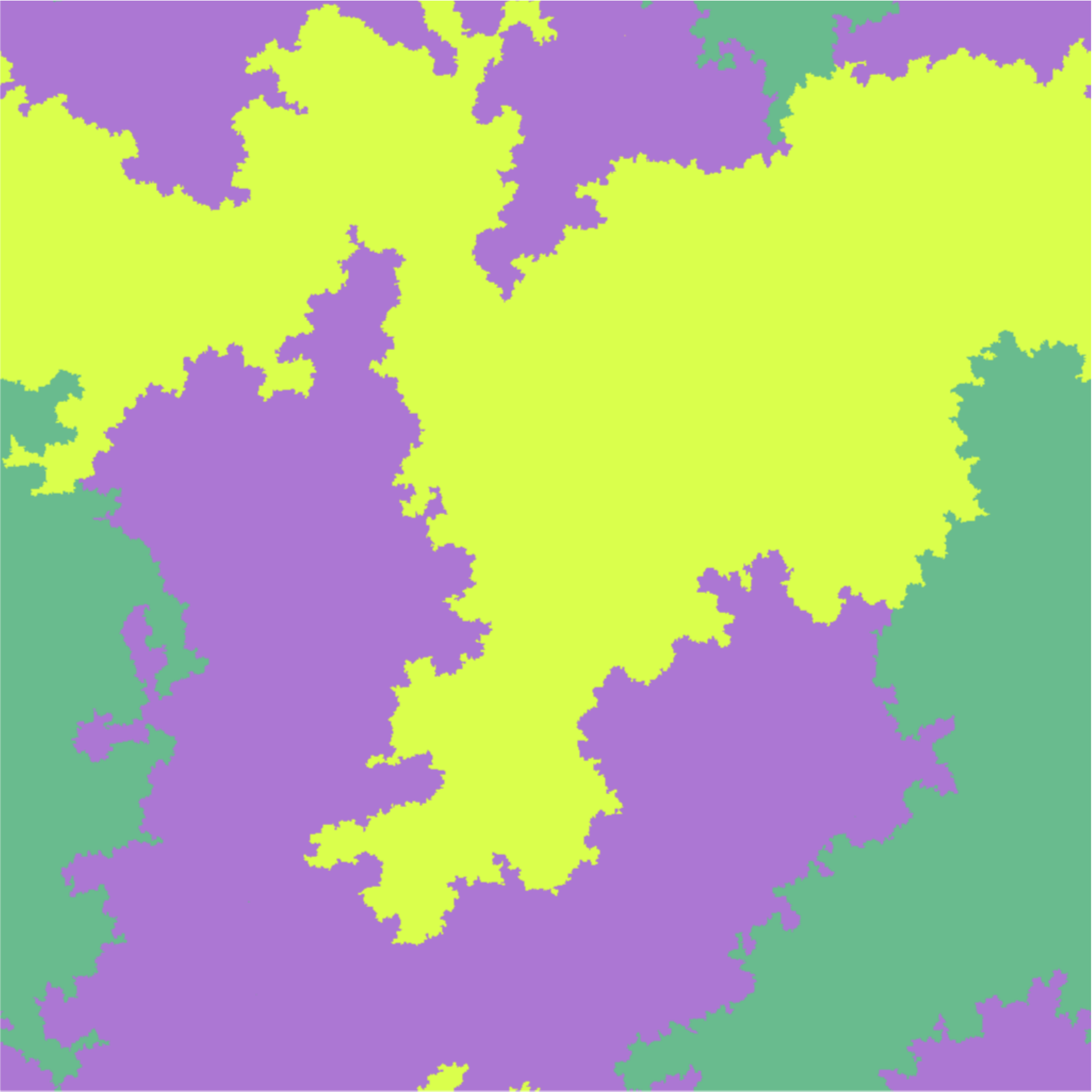}& \includegraphics[width=0.4\textwidth]{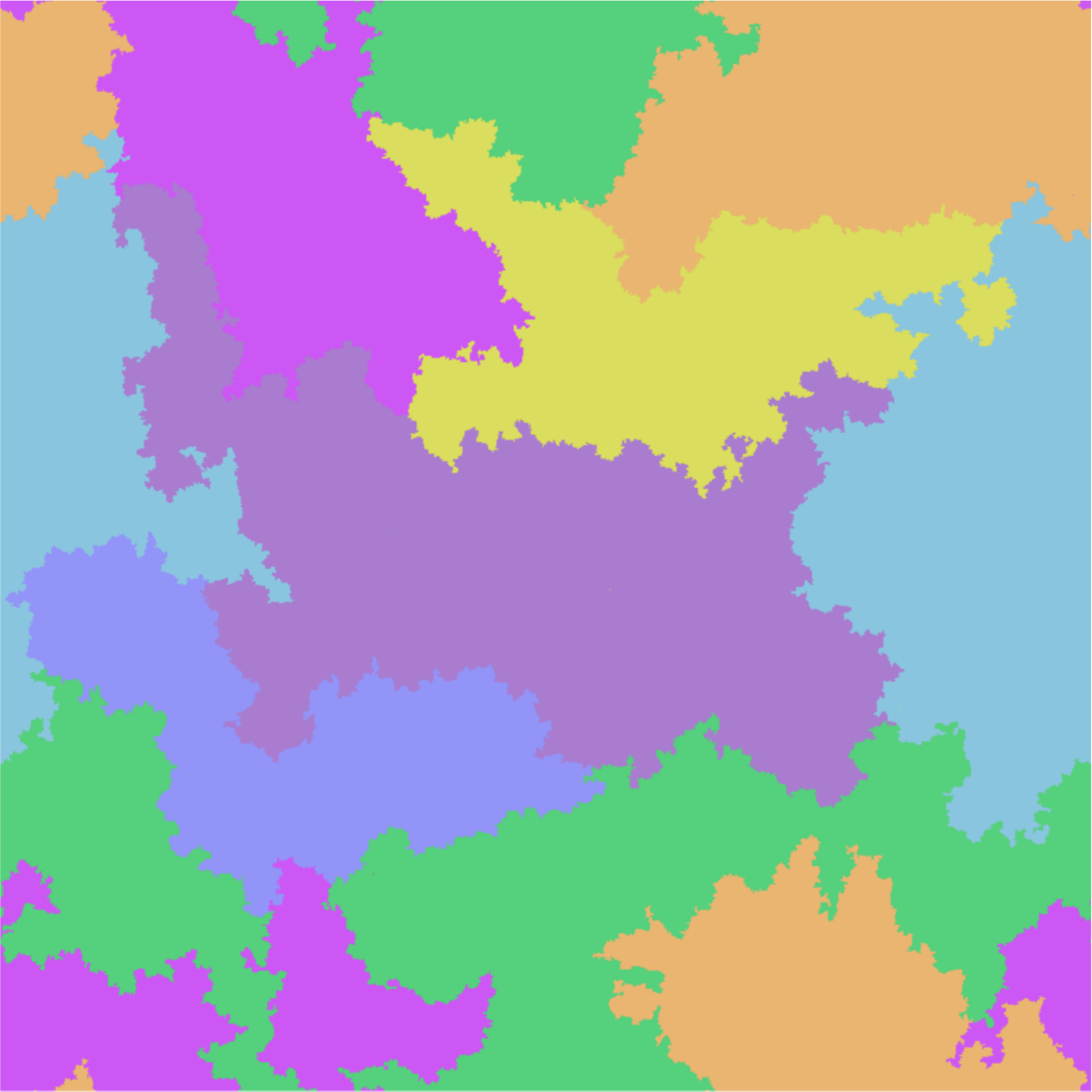}\\
  \end{tabular}
\end{center}
\caption{
  Typical configuration of the system, at the percolation threshold, for the \textit{Largest cluster} and the \textit{Gaussian} models, both with $\alpha=1$.
  Pictures obtained from simulations on a square lattice with $L^2$ sites and $L=1024$.
  Both models yield clear discontinuous transitions and their clusters are compact with the same fractal dimension of the watershed line.
  \label{fig::snapshots} 
}
\end{figure*}
  More recently, the procedure proposed by Achlioptas {\it et al.} has been generalized to a {\it best-of-$m$} product rule in random graphs \cite{daCosta10,Nagler11} and regular lattices \cite{Andrade10} to study its percolation and transport properties.
  The larger the set of bonds, $m$, the lower the probability that the occupied bond is related to the largest cluster, promoting the compactness of the percolation cluster, delaying the percolation threshold, and above an intermediate value, improving the conductivity of the system \cite{Andrade10}.
  Andrade {\it et al.} \cite{Andrade10} have shown that, at the percolation threshold, all exponents for the size dependence of the spanning cluster, the conducting backbone, the cutting bonds, and the global conductance of the system, change continuously and significantly with $m$.

  A \textit{hybrid} model has also been proposed \cite{Araujo11} where an additional parameter is included to interpolate between the discontinuous transition, observed for $m=10$, and the continuous one of classical percolation.
  The model, discloses a nonequilibrium tricritical percolation where explosive percolation is diluted with classical percolation.
  In the diagram for the model two transition lines were identified: a discontinuous and a critical line; both meeting at a tricritical point.
  In the work, the multicritical behavior is characterized by a new set of critical exponents and a tricritical crossover between the discontinuous and the continuous regime is presented.

  Potential applications of explosive percolation are, for instance, the growth dynamics of the Human Protein Homology Network \cite{Rozenfeld10} and the identification of communities in real systems \cite{Pan10}.

  In this manuscript, we start with a description of the two models (\textit{largest cluster} and \textit{Gaussian}) in the next section.
  The nature of the transition and the fractal dimension of the cluster perimeter is discussed in Sec.~\ref{sec::res}, with some final remarks in Sec.~\ref{sec::final}.

\section{Model}\label{sec::model}

In the \textit{largest cluster} model \cite{Araujo10}, bonds are randomly selected from the list of available ones.
If, once occupied, the chosen bond would not lead to the formation or growth of the largest cluster, it is occupied; otherwise, the occupation occurs with probability,

\begin{equation}\label{eq::occuprob}
  \mbox{min}\left\{1,\exp\left[-\alpha\left(\frac{s-\bar{s}}{s}\right)^2\right]\right\} \ \ ,
\end{equation}

\noindent where $s$ is the size of the cluster obtained by occupying the bond, $\bar{s}$ is the average cluster size after the occupation, and $\alpha$ is a parameter of the model that, for simplicity, we take equal to unity.
With this parameter $\alpha$ it is possible to control the size distribution of the clusters.
The larger the value of $\alpha$, the lower the cluster-size dispersion.
For $\alpha\leq0$, all selected bonds are occupied and the model boils down to the classical bond percolation problem, with a continuous transition at the percolation threshold (see, for example, Ref.~\cite{Stauffer94}).
For $\alpha>0$, the formation of a largest cluster differing significantly in size from the average cluster size is systematically demoted, promoting the homogenization of the clusters size.
This homogenization, induces the formation of a ``powder keg'' \cite{Friedman09,Hooyberghs11} which merges at the percolation threshold leading to a discontinuous transition.

The \textit{Gaussian} model \cite{Araujo10} is generalization of the \textit{largest cluster} one.
While in the latter an occupation probability is solely defined to the bonds related with the largest cluster, and all the others are occupied with probability one, in the former, all bonds are occupied with a probability given by Eq.~(\ref{eq::occuprob}).
At each iteration, an unoccupied bond is randomly selected and occupied with this probability, which allows to explicitly control the cluster size distribution.
We denote this model as \textit{Gaussian} since the proposed expression corresponds to a Gaussian distribution with average size $\bar{s}$ and size dispersion $\bar{s}/\sqrt{2}$.
In principle, any function constraining the formation of clusters differing significantly in size from the average cluster size could be considered leading to a discontinuous transition.
Note that, in this model all clusters size are controlled while, in the \textit{largest cluster} model solely the largest cluster is directly controlled and all the smaller ones can freely grow.

Figure~\ref{fig::snapshots} has snapshots of typical configurations, at the percolation threshold, of both models.
Obtained clusters are compact with fractal perimeters.
In the next section, we discuss the discontinuous nature of the transition and the fractal dimension of the largest-cluster interface.

\section{Results}\label{sec::res}

\begin{figure}
\begin{center}
  \includegraphics[width=0.5\textwidth]{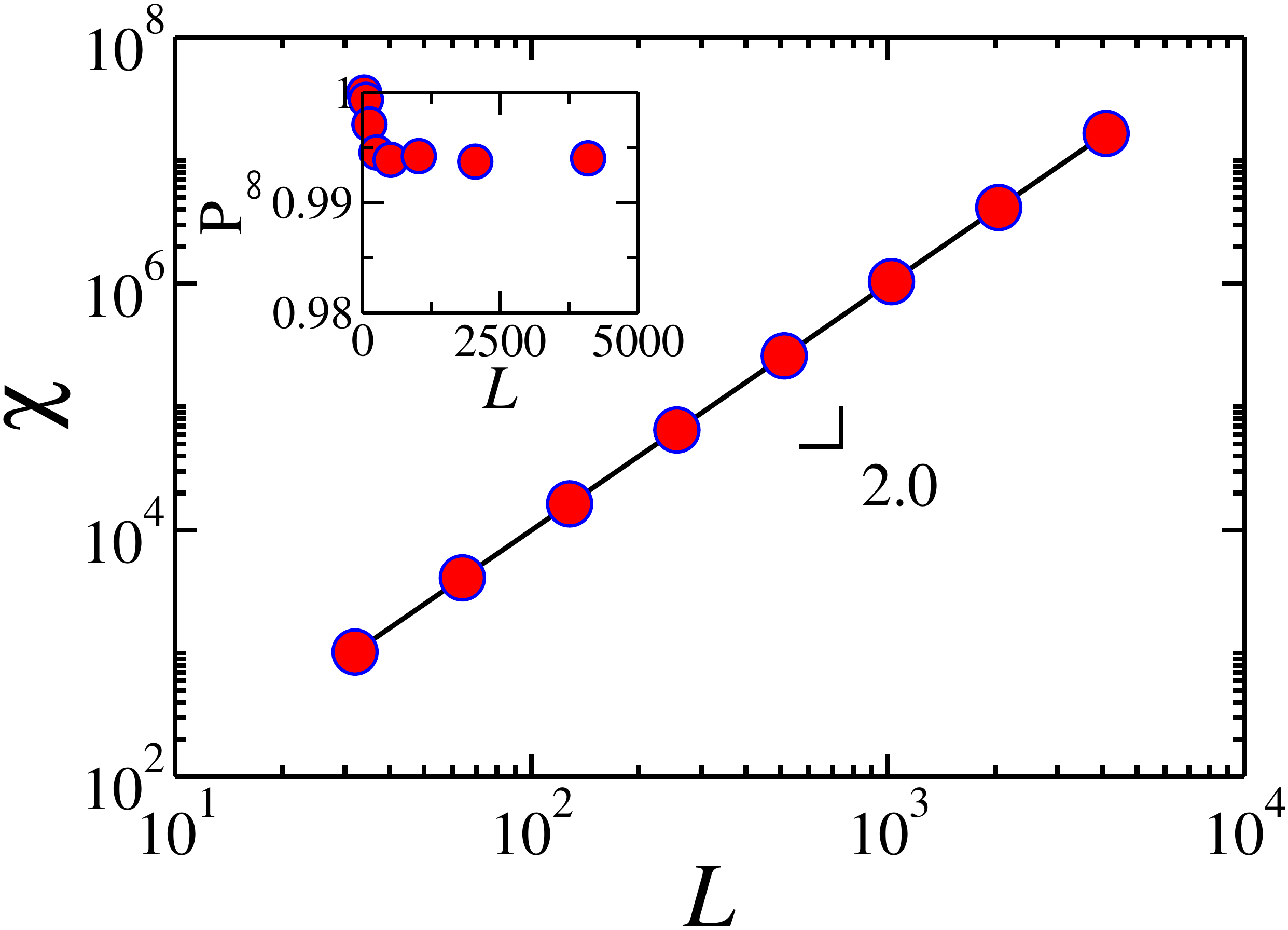}
\end{center}
\caption{
  Size dependence, at the percolation threshold, of the \textit{largest cluster} model, of the susceptibility ($\chi$) and the order parameter ($P_\infty$) -- fraction of sites belonging to the largest cluster. 
  Results have been averaged over $10^4$ realizations of square lattices with $L^2$ sites.
  The linear system size $L$ ranges from $32$ to $4096$.
  The susceptibility scales linearly with the system size and the order parameter converges to a non-zero value in the thermodynamic limit.
  \label{fig::largest} 
}
\end{figure}

  For nonequilibrium problems, where a free energy cannot be defined, transitions can still be classified based on the behavior of the order parameter \cite{Odor04}.
  A discontinuous transition, is characterized by a jump in the order parameter, otherwise, a transition is denoted as continuous.
  For percolation, we define as order parameter the fraction of sites in the largest cluster ($P_\infty$) \cite{Stauffer94}.
  Here we also consider the second moment of the cluster size distribution ($\chi$), defined as

    \begin{equation}\label{eq::sec.mom}
      \chi= \frac{1}{N}\sum_{i} s^2_i \ \ ,
    \end{equation}

  \noindent where the sum runs over all clusters $i$.
  To estimate the percolation threshold we consider the average value of $p$ (fraction of occupied bonds) at which a connected path linking opposite boundaries of the system is obtained.
  Considering different system sizes, for $\alpha=1$, we obtain for the percolation threshold, of the \textit{largest cluster} model, $p_c=0.632\pm0.002$.
  To identify the order of the transition, Fig.~\ref{fig::largest} presents a finite-size study for $P_\infty$ and $\chi$, averaged over $10^4$ samples of square lattices with linear sizes ranging from $32$ to $4096$.
  As we can see in the inset of Fig.~\ref{fig::largest}, above a certain system size, the order parameter, at the percolation threshold, does not show any finite-size dependence, staying at a constant value in the thermodynamic limit ($L\rightarrow\infty$).
  The second moment of the cluster size distribution ($\chi$) scales with $L^d (d=2)$ which is a sign of a discontinuous transition \cite{Binder81,Binder84}.
  For the present models, the percolation thresholds are larger than the ones from previous models due to the compactness of the clusters (see Fig.~\ref{fig::snapshots}).

\begin{figure}
\begin{center}
  \includegraphics[width=0.5\textwidth]{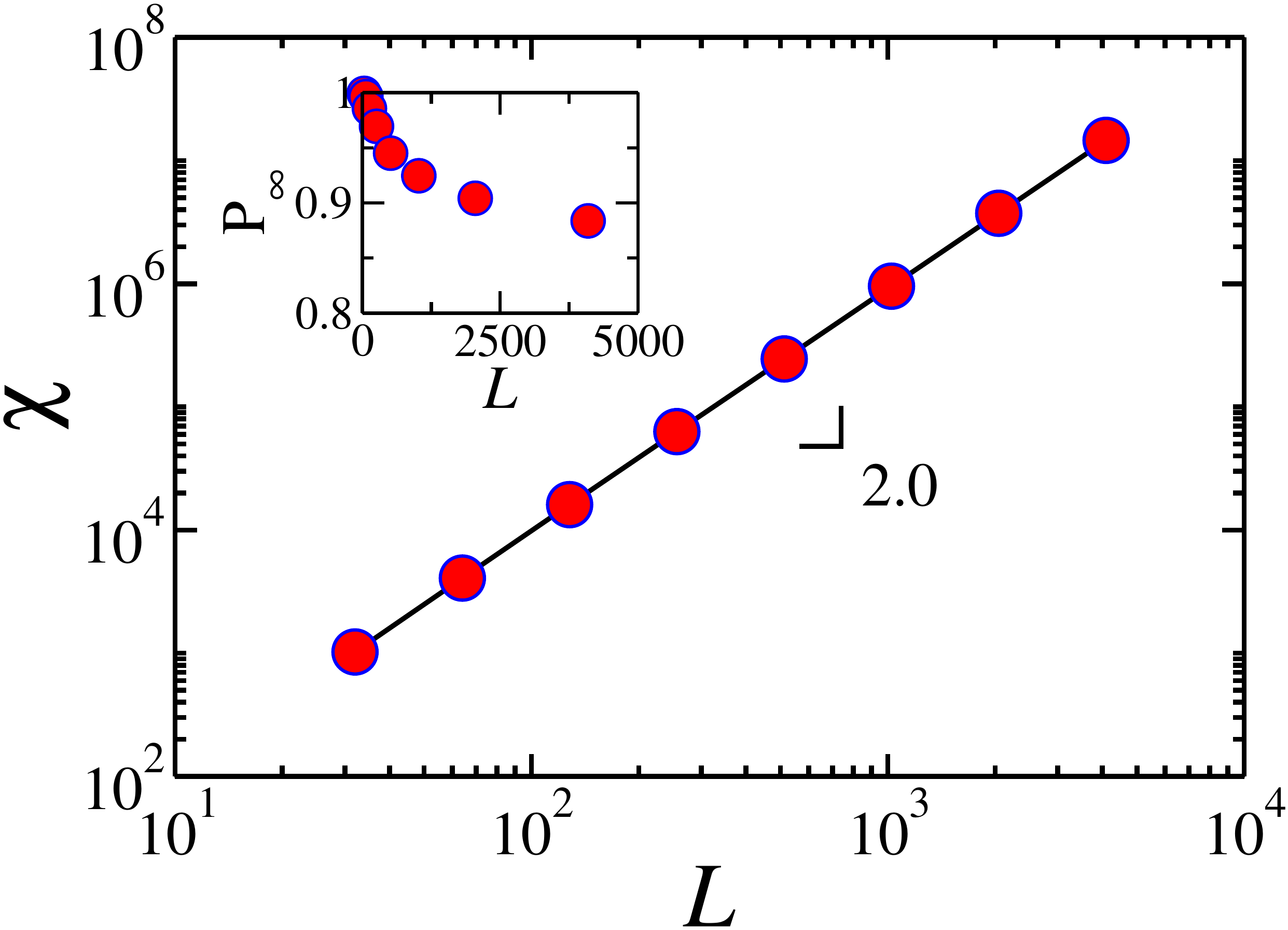}
\end{center}
\caption{
  Size dependence, at the percolation threshold, of the \textit{Gaussian} model, of the susceptibility ($\chi$) and the order parameter ($P_\infty$) -- fraction of sites belonging to the largest cluster. 
  Results have been averaged over $10^4$ realizations of square lattices with $L^2$ sites.
  The linear system size $L$ ranges from $32$ to $4096$.
  The susceptibility scales linearly with the system size and the order parameter converges to a non-zero value in the thermodynamic limit.
  \label{fig::gaussian} 
}
\end{figure}

  As example, for positive $\alpha$, we present, in Fig.~\ref{fig::gaussian}, a size dependence study of the order parameter and second moment of the cluster size distribution, for the {\it Gaussian} model, with $\alpha=1$, at the percolation threshold, on a regular square lattice with linear size ($L$) ranging from $32$ to $4096$.
  Results were averaged over $10^4$ samples.
  We extrapolate, for the infinite system, a percolation threshold $p_c=0.56244\pm0.00006$.
  As for the {\it largest cluster} model, the density of the infinite cluster does not change significantly with the system size and the second moment of the cluster size distribution scales with $L^d (d=2)$.
  As before, these results imply a discontinuous transition.

\begin{figure}
\begin{center}
  \includegraphics[width=0.5\textwidth]{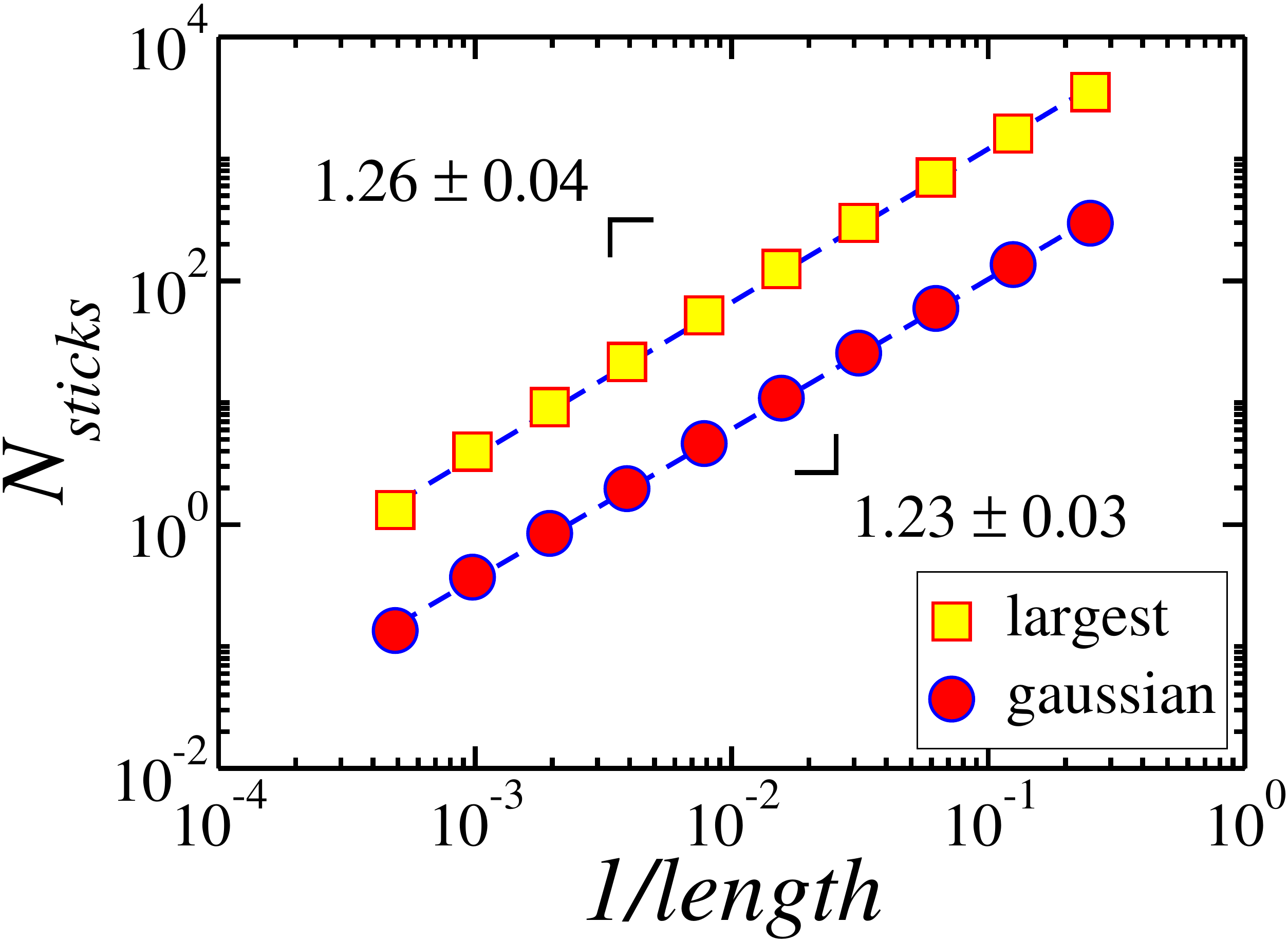}
\end{center}
\caption{
  For the \textit{largest cluster} and the \textit{Gaussian} models ($\alpha=1$), the dependence on the stick size of the number of sticks necessary to follow the perimeter of the infinite cluster, with the yardstick method.
  Results have been averaged over $10^4$ realizations on a square lattice with $2048^2$ sites.
  For visual clarity, data for the \textit{Gaussian} model have been vertically shifted by a factor of $0.1$.
  \label{fig::fracdim} 
}
\end{figure}

\begin{figure}
\begin{center}
  \includegraphics[width=0.5\textwidth]{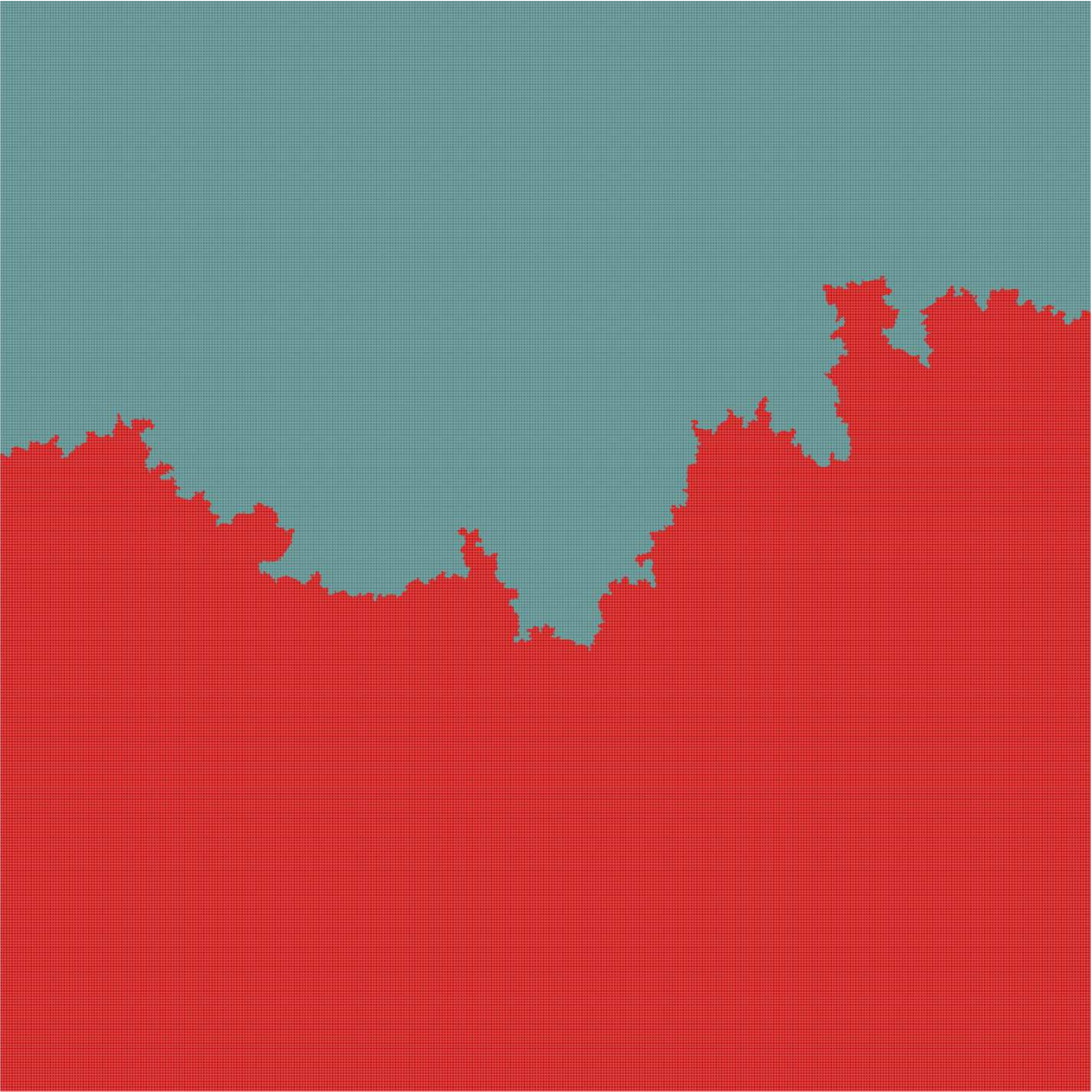}
\end{center}
\caption{
   Snapshot of the watershed line obtained by randomly occupying all bonds in the system except the ones leading to an infinite connection, i.e., closing a path between the bottom and the top of the system.
   The fractal dimension of the interface line is the same as the one of the surface of the infinite cluster of the \textit{largest cluster} and \textit{Gaussian} models.
  \label{fig::watershed} 
}
\end{figure}

  As clearly seen in the snapshots of Fig.~\ref{fig::snapshots}, clusters obtained with our models are compact but we find that the surface is fractal.
  For the {\it Gaussian} model, we calculate for the cluster perimeter a fractal dimension of $1.23\pm0.03$, obtained with the {\it yardstick method} \cite{Tricot88} (Fig.~\ref{fig::fracdim}).
  For the {\it largest cluster} model, it is also characterized by a fractal perimeter with a fractal dimension of $1.26\pm0.04$ (Fig.~\ref{fig::fracdim}).
  Compact clusters with fractal surface were also reported for irreversible aggregation growth in the limit of high concentration by Kolb {\it et al.} \cite{Kolb87}.
  The value of this fractal dimension of percolation is intriguingly close to the one found for watersheds ($1.211\pm0.001$) \cite{Fehr09,Fehr11}, random polymers in strongly disordered media ($1.22\pm0.02$) \cite{Porto99}, and several other models \cite{Andrade11}.
  The simplest way to obtain the watershed line has been proposed by Cieplak \textit{et al.} \cite{Cieplak94} and consists in randomly occupying bonds in the system by systematically suppressing the formation of a path connecting opposite borders of the system, i.e., any bond which leads to the formation of a cluster of connected sites touching the bottom and the top of the system is never occupied.
  As seen in Fig.~\ref{fig::watershed}, in the limit where all the other bonds are selected, only two clusters exist, separated by a watershed line.
  This line is fractal, with the same fractal dimension of the largest-cluster interface of the two models discussed here.
  
\begin{figure}
\begin{center}
  \includegraphics[width=0.5\textwidth]{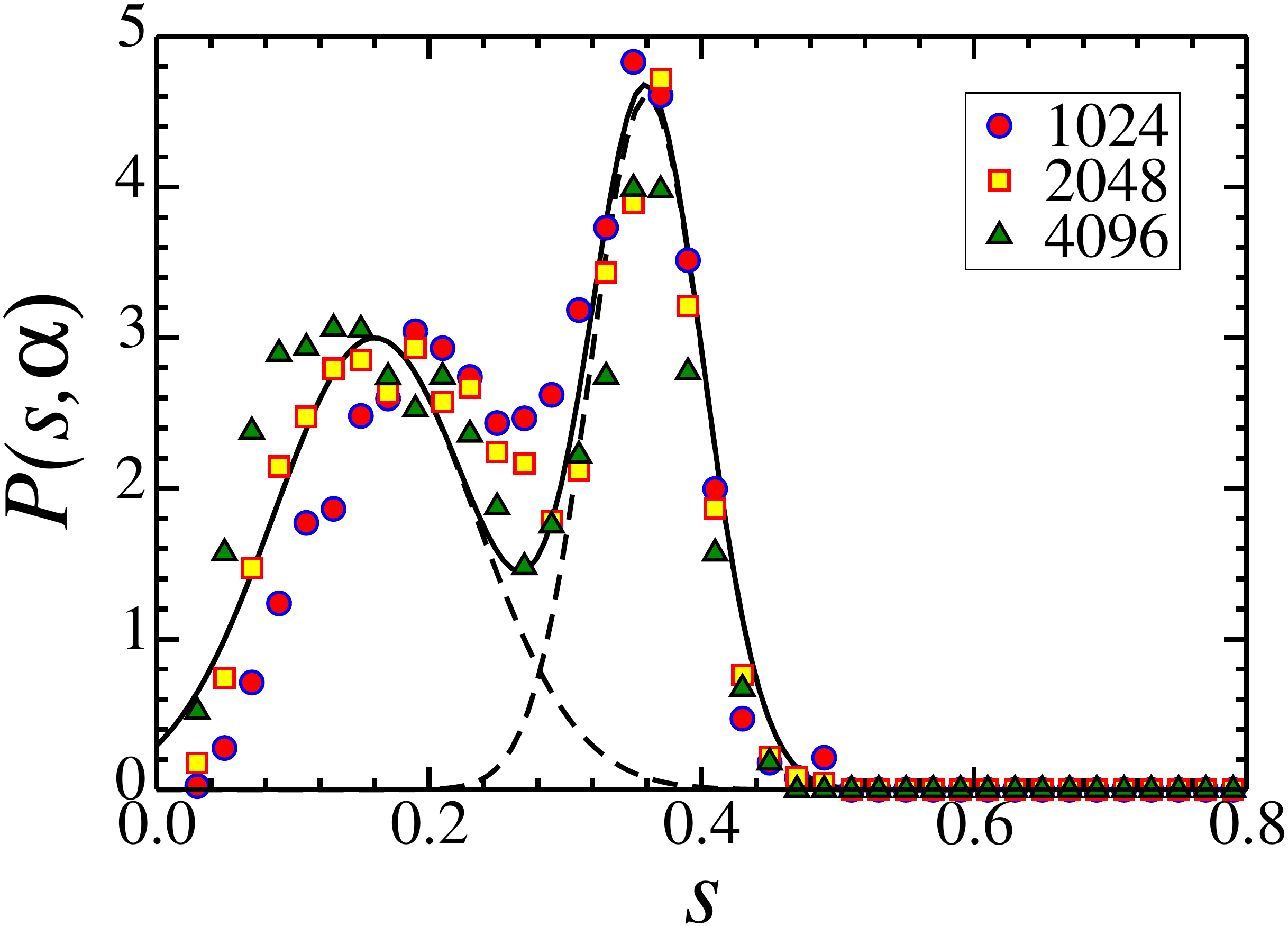}
\end{center}
\caption{
  Cluster size distribution for the percolation threshold of the \textit{Gaussian} model ($\alpha=1$), on a square lattice with $L^2$ sites.
  Results have been averaged over $10^4$ samples and $L=\{1024,2048,4096\}$.
  To reduce finite-size effects, the contribution of the largest clusters has been neglected.
  Black-dashed lines are two Gaussian distributions fitting the results from simulation.
  The black-solid line is the sum of both curves.
  \label{fig::sizedist} 
}
\end{figure}
  
 Figure~\ref{fig::sizedist} shows the cluster size distribution, $P(s,\alpha)$, for different system sizes, obtained with the {\it Gaussian} model.
  Measurements have been performed at the percolation threshold on a square lattice with $1024^2$, $2048^2$, and $4096^2$ sites, and averaged over $10^4$ samples.
  To reduce finite-size effects, we neglect the contribution of the largest cluster.
  Two characteristic peaks are observed, a typical feature of a discontinuous transition.
  For a finite system, at the percolation threshold of such transitions, coexistence of the percolative and non-percolative states is expected \cite{Binder92}.

\section{Final Remarks}\label{sec::final}

  In this manuscript we summarize the main properties of two models of explosive percolation yielding clear discontinuous transitions: the \textit{largest cluster} and the \textit{Gaussian}.
  With the \textit{largest cluster} model we conclude that, on a regular lattice, to obtain an explosive transition is solely necessary to suppress the growth of the largest cluster what, indirectly, promotes the homogenization of the clusters size.
  With the \textit{Gaussian} model the discontinuous nature of the transition is supported by a bimodal cluster-size distribution resulting from the coexistence of a percolative and non-percolative state.
  For both models, the fractal perimeter of the largest cluster is intriguingly close to the one found for watersheds ($1.211\pm0.001$) \cite{Fehr09}, random polymers in strongly disordered media ($1.22\pm0.02$) \cite{Porto99}, optimal path cracking \cite{Andrade09,Oliveira11}, and several other models \cite{Andrade11}, and we have arguments that they are actually identical.

\section*{Acknowledgments}
We acknowledge financial support from the ETH Competence Center Coping with Crises in Complex Socio-Economic Systems (CCSS) through ETH Research Grant CH1-01-08-2.
We also acknowledge the Brazilian agencies CNPq, CAPES and FUNCAP, and the grant CNPq/FUNCAP, for financial support.

\bibliography{text.bib}

\end{document}